\documentclass[aps,prd,superscriptaddress,twocolumn,preprintnumbers,nofootinbib,endnote]{revtex4-1}

\pdfoutput=1

\usepackage{amsfonts,amssymb,amsmath}
\usepackage{epsfig}

\newcommand{\SU}{{\rm SU}}

\newcommand{\jets}{\text{jets}}
\newcommand{\jet}{\text{jet}}

\newcommand{\ra}{\rightarrow}

\newcommand{\tev}{~\text{TeV}}
\newcommand{\gev}{~\text{GeV}}

\newcommand{\pb}{~\text{pb}}
\newcommand{\fb}{~\text{fb}}

\newcommand{ \slashchar }[1]{\setbox0=\hbox{$#1$}   % set a box for #1
   \dimen0=\wd0                                     % and get its size
   \setbox1=\hbox{/} \dimen1=\wd1                   % get size of /
   \ifdim\dimen0>\dimen1                            % #1 is bigger
      \rlap{\hbox to \dimen0{\hfil/\hfil}}          % so center / in box
      #1                                            % and print #1
   \else                                            % / is bigger
      \rlap{\hbox to \dimen1{\hfil$#1$\hfil}}       % so center #1
      /                                             % and print /
   \fi}                                             %

\begin{document}
%%%%%%%%%%%%%%%%
%%%%%%%%%%%%%%%%

\preprint{FERMILAB-PUB-10-509-PPD-T}

%\title{Higgs discovery through vector-like quarks using jet substructure}
\title{Higgs Discovery through Top-Partners using Jet Substructure}

\author{Graham D. Kribs}
\affiliation{Theoretical Physics Department, Fermilab, Batavia, IL 60510}
\affiliation{Department of Physics, University of Oregon, Eugene, OR
  97403}

\author{Adam Martin}
\affiliation{Theoretical Physics Department, Fermilab, Batavia, IL 60510}

\author{Tuhin S. Roy}
\affiliation{Department of Physics, University of Washington, Seattle, WA
  98195} 

\begin{abstract}

Top-partners -- vector-like quarks which mix predominantly with the 
top quark -- are  simple extensions of the standard model present 
in many theories of new physics such as little Higgs models, 
topcolor models, and extra dimensions.
Through renormalizable mixing with the top quark, these top-partners 
inherit  couplings to the Higgs boson.  Higgs bosons produced from 
the decay of top-partners are often highly boosted and ideal candidates 
for analyses based on jet substructure.  Using substructure methods, 
we show that light Higgs bosons decaying to $\bar b b$ can be discovered 
at the $14\tev$ LHC with less than $10\fb^{-1}$ for top-partner masses 
up to $1\tev$.

\end{abstract}

\maketitle

%%%%%%%%%%%%%%%%%%%%%%
\section{Introduction}
%%%%%%%%%%%%%%%%%%%%%%

While the fermions of the standard model (SM) are notoriously chiral, 
extensions of the SM often involve new fermions whose left and 
right-handed components have the same quantum numbers, so-called 
vector-like fermions. Because of this charge assignment, vector-like 
fermions do not require electroweak symmetry breaking to have mass, 
making them a particularly self-contained, worry free extension 
of the SM. The interactions between new vector-like fermions and the 
matter in the SM depend on the quantum numbers of the new fermions. 
One common choice is to include new fermions with the same 
quantum numbers as the right-handed quarks and which mix 
predominantly with the top quark: a top-partner. 
The restriction to mixing with the top quark is driven in part 
by necessity -- the interactions of lighter fermions are more 
tightly constrained by experiment -- and in part by the idea 
that the large mass of the top is an indication that it is more 
intimately connected with new ultraviolet physics.  Despite their vector-like 
nature, these top-partners do interact with the Higgs boson because of 
mass mixing. The goal of this paper is to demonstrate that this 
coupling between vector-like top-partners and the Higgs boson 
can have profound impact on the discovery potential of a light 
Higgs boson at the LHC.

While certainly phenomenologically interesting, extending the SM with a vector-like top-partner is also well-motivated from a theoretical perspective. In order to maintain a naturally light Higgs boson, divergent quantum corrections from loops of top quarks must be removed. One way to reduce the top-induced divergence is to enlarge the approximate global symmetries of SM, which requires new particles, including top-partners. In little Higgs theories~\cite{ArkaniHamed:2001nc,ArkaniHamed:2002qx,ArkaniHamed:2002qy,Low:2002ws,Schmaltz:2002wx,Kaplan:2003uc,Skiba:2003yf,Chang:2003un,Chang:2003zn,Cheng:2003ju,Katz:2003sn,Cheng:2004yc,Schmaltz:2004de,Low:2004xc}, the top-loop divergence is canceled precisely by a loop of a vector-like top-partner with exactly the properties described above.  Some of the models are sufficiently safe from electroweak precision constraints \cite{Csaki:2002qg,Hewett:2002px,Csaki:2003si,Chen:2003fm} to allow these vector-like top-partners to be accessible at the LHC \cite{Burdman:2002ns,Han:2003wu,Perelstein:2003wd,Cheng:2005as}. Vector-like quarks are also prevalent in extra-dimensional models~\cite{Cheng:1999bg, Carena:2006bn, Contino:2006qr,Burdman:2007sx} as the Kaluza-Klein (KK) excitations of SM fermions, in top-color models~\cite{Hill:1991at}, in supersymmetric little higgs models~\cite{Birkedal:2004xi, Chankowski:2004mq, Chacko:2005ra, Roy:2005hg, Csaki:2005fc, Falkowski:2006qq, Chang:2006ra, Bellazzini:2009ix} and even in the simplest extensions of the SM where gauge coupling unification serves as the only guiding principle besides a WIMP dark matter, e.g.~\cite{Kilic:2010fs}. 

Collider studies of vector-like top-partners are plentiful~\cite{Burdman:2002ns,Han:2003wu,Perelstein:2003wd, Azuelos:2004dm, Cheng:2005as, Karafasoulis:962029, Matsumoto:2006ws, Skiba:2007fw, Holdom:2007nw, Holdom:2007ap, :2008nf, Lister:2008is, AguilarSaavedra:2009es}. However, past searches focus on the discovery of the top-partners, rather than using the top-partners to enhance the discovery of {\em other} particles. Direct collider searches put a lower bound of $\sim$ few hundred GeV on the mass of the top-partners~\cite{:2008nf, Lister:2008is}. New colored particles can be copiously produced at the LHC, and through the top-partner--Higgs interaction, top-partner decays will generate a new source of Higgs bosons.  A new production mode of Higgs bosons is most interesting when Higgs boson is light (namely, $m_h \lesssim 130\gev$).  In this mass range, the Higgs boson requires over $10$~fb$^{-1}$ to find in one of the more traditional SM searches, such as $h\rightarrow \tau^+ \tau^-$ and $h \rightarrow \gamma\gamma$, and over $30$~fb$^{-1}$ to find in the $h \rightarrow \bar{b}b$ channel using jet substructure techniques \cite{Butterworth:2008iy}.

One complication in this scenario is that top-partners tend to interact with, and therefore, decay primarily to the third generation quarks. So, while the decay of top-partners does occasionally yield Higgs bosons, the decay products of the Higgs bosons are surrounded by a sea of top and bottom quarks. The large number of bottom quarks is especially problematic since it introduces a large combinatoric impediment to Higgs reconstruction via the dominant $h\rightarrow \bar b b$ mode. 

In this paper we look for \emph{boosted} light Higgs bosons that decay through the $h\rightarrow \bar b b$ mode. Boosted Higgs boson decay to $b\bar{b}$ can be captured by looking for jet with substructure consistent with massive particle decay. The shift to large jets and boosted objects has several advantages: (i) the high mass, distinct substructure, and heavy flavor content within the large jets are all effective handles which can be used to separate the signal from the background~\cite{Butterworth:2002tt, Butterworth:2008iy, Thaler:2008ju, Brooijmans:2008se, Almeida:2008yp, Almeida:2008tp, Butterworth:2009qa, Ellis:2009su, Ellis:2009me, Plehn:2009rk, Krohn:2009zg, Krohn:2009th, Chekanov:2010vc, Soper:2010xk, Hackstein:2010wk, Katz:2010mr}; and (ii) by capturing the Higgs boson entirely within a single jet we can reduce the combinatorial problem -- only $b$\,-jets which are close enough to be encompassed by a single fat jet are used for resonance reconstruction.

Previous attempts have shown that Higgs boson can be discovered spectacularly well in the supersymmetric extension of the SM using jet substructure techniques even from a sample of $10\fb^{-1}$ of data~\cite{Kribs:2009yh, Kribs:2010hp}.  What sets this work apart is the fact that unlike weak scale supersymmetry,  the final states consist solely of SM particles and hence do not automatically have large missing energies.  Without a clean sample of new physics events, finding Higgs resonance is more challenging. We propose a strategy that combines various boosted object taggers (such as top and $W/Z$ taggers) with conventional cuts and requirements.  We find that for top-partners up to about $800\gev$, the algorithm is capable of discovering the Higgs boson in the $b\bar{b}$ channel with high significance before any SM search.

The setup of this paper is the following: In Sec.~\ref{sec:setup} we describe the minimal vector-like top-partner model and define the mass eigenstates and couplings. Next, in Sec.~\ref{sect:phenostart} we describe the analysis strategy -- the tools, both conventional and unconventional that we will use, the flow of analysis cuts. In Sec.~\ref{sect:sim} we give further simulation details and present our results. Following our main results, we provide two example models whose low energy effective theory contains the exact states and interactions necessary for our study (Sec.~\ref{sec:models}). We then conclude in Sec.~\ref{sec:conclude} with a discussion.

\section{Mixing top-partners with the top}
\label{sec:setup}
 
While there are many possible vector-like extensions of the SM, in this work we will focus on the vector-like top. We enlarge the SM by two Weyl fermions: $T \equiv \left( 3 , 1 \right)_\frac{2}{3}\, ,\,T^c  \equiv \left( \bar{3} , 1 \right)_{-\frac{2}{3}}$. With these quantum numbers, the simplest, renormalizable interactions we can write down involving the new fermions are:

\begin{equation}
  \label{eq:1}
    \mathcal{L} \supset y_1   Q_3 H t^c + \delta\, T  t^c  +  M T   T^c, 
 \end{equation}
where $Q_3$ is the (SM) third generation electroweak doublet ($t, b$). Note that an additional term $ Q_3 H T^c $ is also allowed under all symmetries. It can, however, be eliminated by rotating $t^c$ and $T^c$ and consequently redefining $\delta$ and $M$.
 
The mass eigenstates are combinations of the quarks $t^c$ and $T^c$. The full mass matrix, including nonzero Higgs vacuum expectation value (vev) is given by 
\begin{equation}
  \label{eq:2}
    \begin{pmatrix}          t  & T    \end{pmatrix} 
    \begin{pmatrix}
                m     &   0 \\
               \delta & M 
   \end{pmatrix}
   \begin{pmatrix}     t^c \\ T^c   \end{pmatrix}  \; ,
\end{equation}
where $ m = \frac{y_1}{\sqrt 2}\, v$ and $v$ is the Higgs vev. In general, this mass matrix is not symmetric and can be diagonalized by a bi-unitary transformation which rotates the left handed and the right handed quarks by different angles. Under such a rotation, the quark mass eigenstates are related to the quarks in Eq.~(\ref{eq:1}) in the following way
\begin{equation}
\begin{split}
  \label{eq:3}
  \begin{pmatrix}     t_1 \\ t_2  \end{pmatrix} 
    & = \begin{pmatrix} 
      \cos \theta_l & - \sin \theta_l \\
      \sin \theta_l & \cos \theta_l 
       \end{pmatrix} 
     \begin{pmatrix}     t  \\ T    \end{pmatrix}  
\\ 
   \begin{pmatrix}     t_1^c \\ t_2^c   \end{pmatrix} 
    & = \begin{pmatrix} 
      \cos \theta_r & - \sin \theta_r \\
       \sin \theta_r & \cos \theta_r 
       \end{pmatrix} 
     \begin{pmatrix}     t^c \\ T^c   \end{pmatrix}  .
\end{split}
\end{equation}
The angles $\theta_l$, $\theta_r$  and the mass eigenvalues can be determined to be
\begin{align}
  \label{eq:5}
  &  \qquad \theta_l = \frac{1}{2} \tan^{-1} \left( 
                 \frac{  2\, \delta\, m }{M^2 - m^2 + \delta^2} \right) \nonumber \\
  & \qquad \theta_r = \frac{1}{2} \tan^{-1} \left( 
                \frac{   2\, \delta\, M  }{ M^2 - m^2 - \delta^2} \right)  \\
   m_t &= \cos \theta_l \cos \theta_r \left( m + \delta\, \tan \theta_l + 
                   M\, \tan \theta_l \tan \theta_r  \right) \nonumber \\
m_T &= \cos \theta_l \cos \theta_r \left( M - \delta\, \tan \theta_r + 
                   m \tan \theta_l \tan \theta_r  \right) \nonumber
\end{align}
In a more useful parametrization, the angles and the couplings may be expressed as a function of $m_t, m_T$ and $\eta = \delta/M$.  
\begin{align}
  \label{eq:13}
  \tan \theta_r & = \frac{m_T}{m_t}  \tan \theta_l \\
  \theta_l & = \frac{1}{2} \sin^{-1} \left(  \eta  \frac{2 m_t m_T}{ m_T^2 - m_t^2} \right)
\end{align}

In order to study the collider phenomenology of this minimal setup we introduce four component Dirac spinors $t_D = \begin{pmatrix} t_1 \\ t_1^{c\dag} \end{pmatrix}$  and  
$T_D = \begin{pmatrix} t_2 \\ t_2^{c\dag} \end{pmatrix}$. The non-diagonal interactions of heavy top can then be recast in the form 
\begin{equation}
  \begin{split}
\mathcal{L} &    \supset \frac{m_t\, \cos^2{\theta_l}}{v} 
              h ~ \bar T_D ( \tan \theta_r\, P_L +  \tan \theta_l\, P_R)\, t_D \\
&  + \frac{g_2 \sin \theta_l\, \cos{\theta_l}}{2\,\cos{\theta_W}} 
          Z_\mu \left( \bar T_D \gamma^{\mu}\,P_L t_D + \bar t_D \gamma^{\mu}\, P_L T_D \right)   \\
 & \  +  \frac{g_2 \sin \theta_l}{\sqrt{2}} 
      \left(W^{+}_\mu \bar T_D \gamma^{\mu}P_L  b_D + W^{-}_\mu \bar b_D \gamma^{\mu}\,P_L T_D \right),
 \end{split}
\label{eq:8}
 \end{equation}
where $P_{L,R}$ are the usual projectors.

\begin{figure}[t]
\centering
\includegraphics[width=0.45\textwidth]{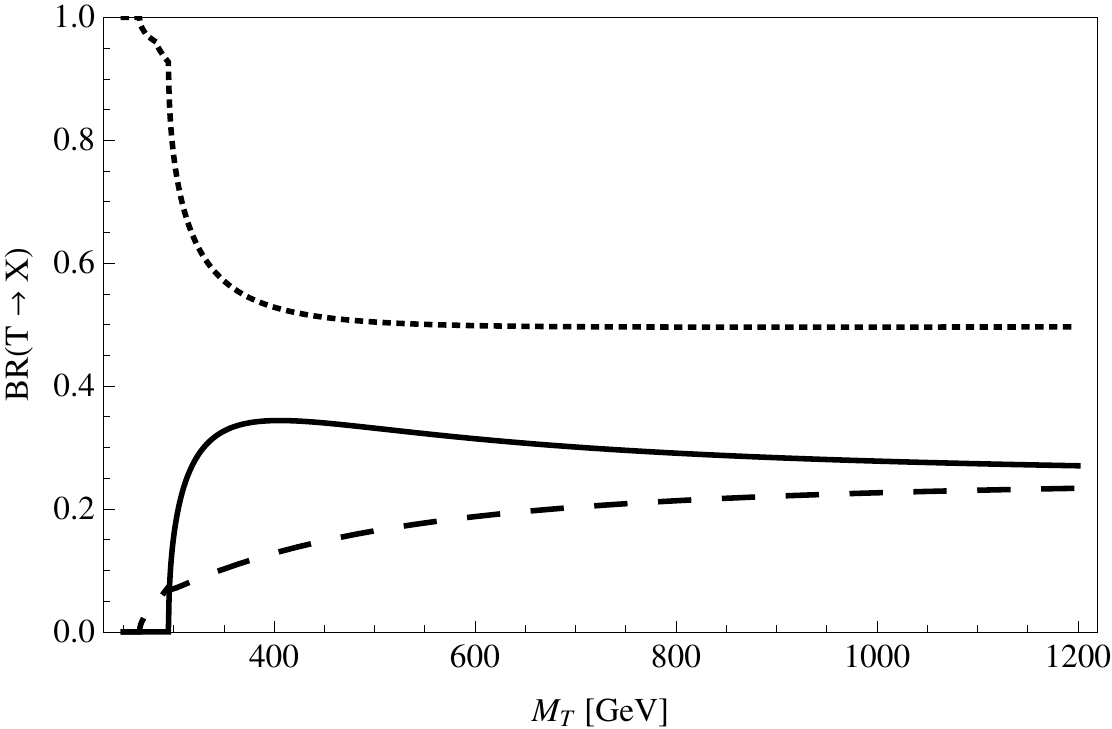}
\caption{Branching fraction of $T$ to $t + h,\ m_h = 120\gev$ (solid), $b + W$ (dotted) and $t + Z$ (dashed) as a function of $m_T$. An $\eta$ value of $0.5$ has been chosen, though the branching ratios are essentially independent of $\eta$. }
\label{fig:bratios}
\end{figure}

Within the minimal model, the top-partner can only decay to the Higgs boson, $W$ or $Z$. In the limit of infinite $T$ mass, the branching fractions for these modes are essentially governed by ``Goldstone equivalence": $T$ can only decay into Higgs degrees of freedom via  the first term in Eq.~(\ref{eq:1})\,-- two of these degrees of freedom are eaten to become the longitudinal polarization of $W^{\pm}$, one is eaten by the $Z$, and the remaining one is the physical Higgs boson. Therefore, we have:
\begin{equation}
\begin{split}
 & BR(T \rightarrow t + h)  \sim 25 \%,  \quad BR(T \rightarrow t + Z)  \sim 25 \%    \\
 & \qquad \qquad \quad BR(T \rightarrow  b + W ) \sim 50 \%. 
\end{split}
\label{eq:6}
\end{equation}
Different kinematics among the three modes alters this ratio, especially for lighter $m_T$, however it remains a decent approximation~\cite{Dobrescu:2009vz}. To demonstrate this, we plot the branching fraction to the three modes as a function of $m_T$ in Fig.~(\ref{fig:bratios}).

\section{Boosted Higgs bosons from top-partners}
\label{sect:phenostart}

\subsection{Pair production versus single production of top-partners}

\begin{figure}[t]
\centering
\includegraphics[width=0.48\textwidth]{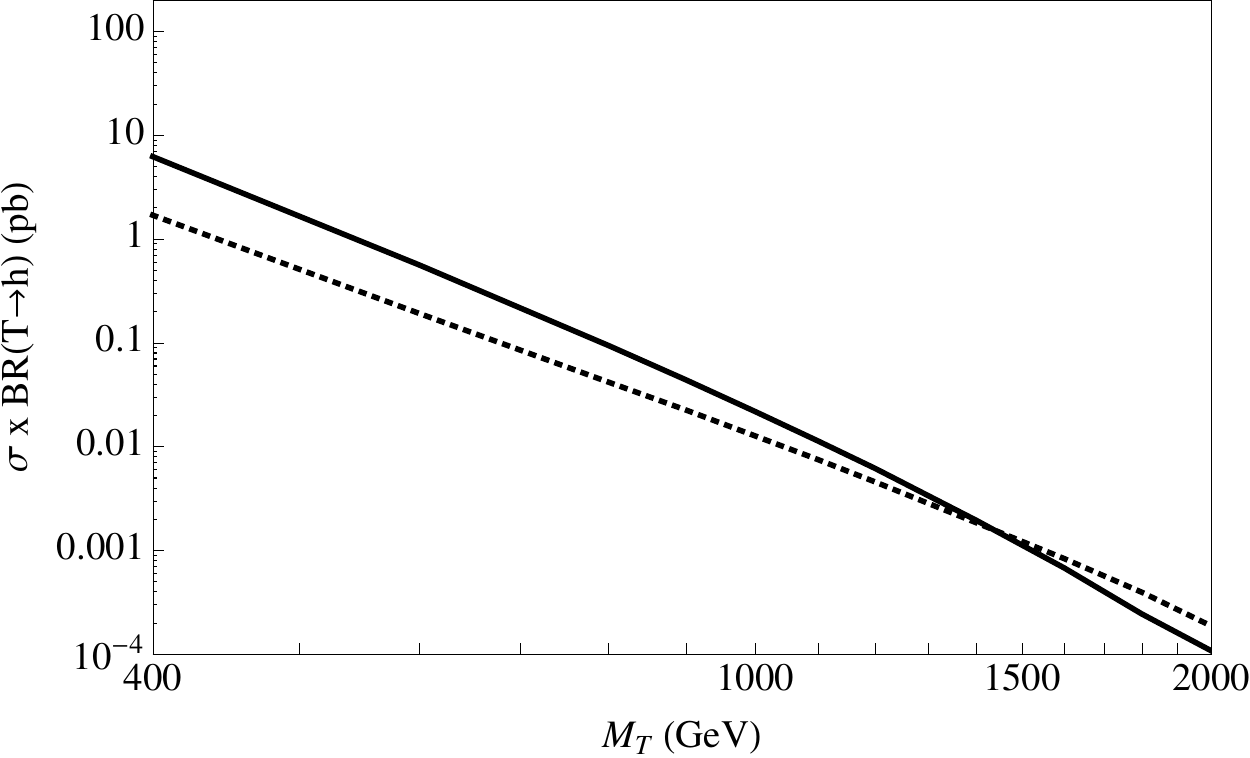}
\caption{Comparison of the production cross sections $\sigma(pp\rightarrow T\bar T)$ (solid) and $\sigma(pp\rightarrow T+j, \bar T + j)$ (dotted) with at least one $T \ra t h$ at a 14 TeV LHC\@. The $\eta$ parameter, which enters into single production, has been set to $1/2$. Smaller $\eta$ decreases the cross section slightly.}
\label{fig:production}
\end{figure}

Because of the $T$-$b\,$-$W$ coupling in Eq.~\eqref{eq:8}, single production of $T$ is possible. However, the cross section depends on the $b$-quark pdf of proton, proportional to the electroweak coupling,  and additionally suppressed because of $W$ exchange in the $T-$channel. As long as $m_T \lesssim 1.1\tev$, single production is always subdominant with respect to the QCD pair production of $T$~\cite{Han:2003wu, Azuelos:2004dm}.  The dominance of the pair production below $1.1\tev$ is demonstrated in Fig.~\ref{fig:production}.  

Single production, while subdominant, does nevertheless create a cleaner final state compared to pair production, for example $pp \rightarrow T + q$. Cleaner states are certainly easier to reconstruct, however one $T$ resonance in the event obviously provides fewer handles for distinguishing signal and background compared to $T$ pair production. We find the cleaner final state does not compensate sufficiently for the lack of handles, so single production is always inferior to pair production, at least for the purpose of Higgs discovery. Therefore, in this work we concentrate on the following set of topologies: pair production of $T$ followed by the decay of one $T$ to $t + h$ and the other $T$ to $b+W$ or $t+Z$.
\begin{figure}[t]
\centering
\includegraphics[width=0.4\textwidth]{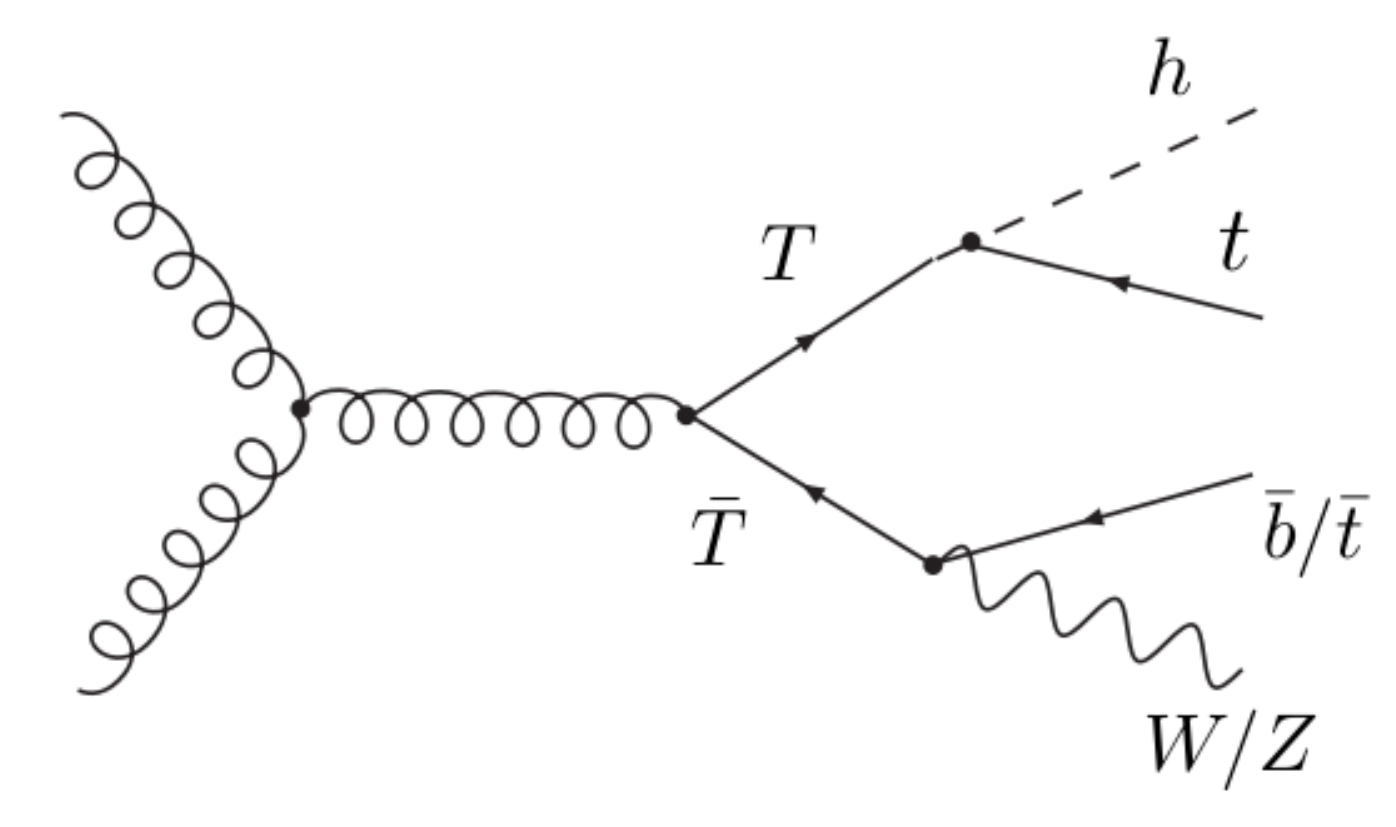}
\caption{A sample Feynman diagram for $T \bar{T}$ pair production followed by decays to a Higgs boson and a $W$ or $Z$.}
\end{figure}

\subsection{Search Strategy}
\label{sect:search}

In order to come up with a successful search strategy, we first need to understand the standard model backgrounds as well as new physics backgrounds that we must overcome. Every interesting signal event contains multiple resonances, meaning Higgs bosons, $W$, $Z$, or tops. More specifically, in addition to the Higgs boson, there is always at least one top quark, one gauge boson and one $b$ quark. Signal $W$ bosons and $b$ quarks can either come directly from the decay of the top-partner, or they can come from the decay of the top. The dominant SM backgrounds are $\bar t t + \jets$, $\bar t t + \bar b b$ and $W/Z + \jets$ -- all processes with large cross section containing gauge bosons and multiple hard jets. We will restrict our search to final states which contain at least one lepton to avoid an overwhelming QCD multi-jet background. The specifics of the backgrounds, including cross sections and generator details, will be given in Sec.~\ref{sect:sim}.  
 
The success of our search for a boosted Higgs boson relies crucially on  combinations of conventional handles (such as existence of isolated leptons and large $H_T$ {\it i.e.} scalar sum of visible energies in an event) and slightly unconventional tools (boosted object taggers). Each of these handles is described in more detail in the following subsection.

\begin{itemize}
\item {\bf isolated lepton:} In our simulation leptons are considered as isolated they have $p_T > 15\ \gev,\ |\eta| < 2.5$  and if the energy deposited by hadrons within a cone of size $R=0.4$ surrounding the lepton is less than $20\%$ of the energy deposited by the lepton. Our simple implementation tags leptons with a $90\%$ efficiency.

\item {\bf $\mathbf{H_T}$:} $H_T$ is defined as the scalar sum of all visible energy in the detector with $|\eta| < 4.0$. We calculate it by summing up the energies of all particles except neutrinos. Also note that after the hadrons are  granularized into calorimeter cells, we disregard all cells with energy less than $1\gev$ and so these do not contribute to our estimation of $H_T$ in an event.

\item {\bf top-tagger:} Top taggers are designed to find a hadronic top when it is boosted enough so that its decay products lie in one single jet. Examples of top taggers can be found in various recent works~\cite{Thaler:2008ju, Kaplan:2008ie, Almeida:2008yp, Almeida:2008tp, Ellis:2009me, Ellis:2009me, Plehn:2009rk, Plehn:2010st}. We employ the top-tagger introduced in Ref.~\cite{Plehn:2009rk} as it has been demonstrated to work significantly better with lower boost~\cite{spannowsky-talk} and in particularly complex environments.  In a sample of $t\bar{t}h$ events the tagger is found to have identified top with $43\%$  efficiency and with $5\%$ mistag rate when used on a sample of $W+\text{jets}$. Following~\cite{Plehn:2009rk}, the inputs to the top-tagger are jets of size $R=1.5$ which have been clustered with the Cambridge/Aachen algorithm~\cite{Dokshitzer:1997in, Wobisch:1998wt, Wobisch:2000dk}. The details of the declustering and subjet identification procedures can be found in Ref.~\cite{Plehn:2009rk}.   

\item {\bf $W/Z$ tagger:}   Like the top-tagger, the $W/Z$ tagger also identifies hadronic and boosted $W/Z$. We use a modified version of an algorithm used by Butterworth et. al. (BDRS)~\cite{Butterworth:2008iy} to find boosted Higgs bosons. The tagger consists of two parts: First, a jet is checked whether it contains both the partons from the decay of a massive particle going through a two body decay ({\it i.e.} the mass of the subjets are significantly smaller than the mass of the jet and the splitting of the jet into the two subjets is not too  ``asymmetric'').  Jets which pass this criteria are ``filtered ''  and retained if the mass of the filtered jet is within $\left(65-95\right)\gev$.  We use Cambridge/Aachen jets but  with smaller size (namely, $R=1.2$) as inputs to the $W/Z$ tagger. We use the same declustering parameters as BDRS, however we \emph{do not} require the subjets to carry heavy flavor. We find high-$p_T$ $W/Z$ events are tagged with $\sim 80\%$ efficiency, while QCD jets with $p_T \gtrsim 200\gev$ are misidentified as $W/Z$ roughly $\sim 10\%$ of the time. The resulting tagged jet mass distribution, when the tagger works on  $W/Z$ samples and for QCD-jets, is shown below in Fig.~\ref{fig:wtagged}.  As expected, the QCD mistagged jets do not show any profile of a resonance in their jetmass distribution.
\begin{figure}[t]
\centering
\includegraphics[width=0.48\textwidth]{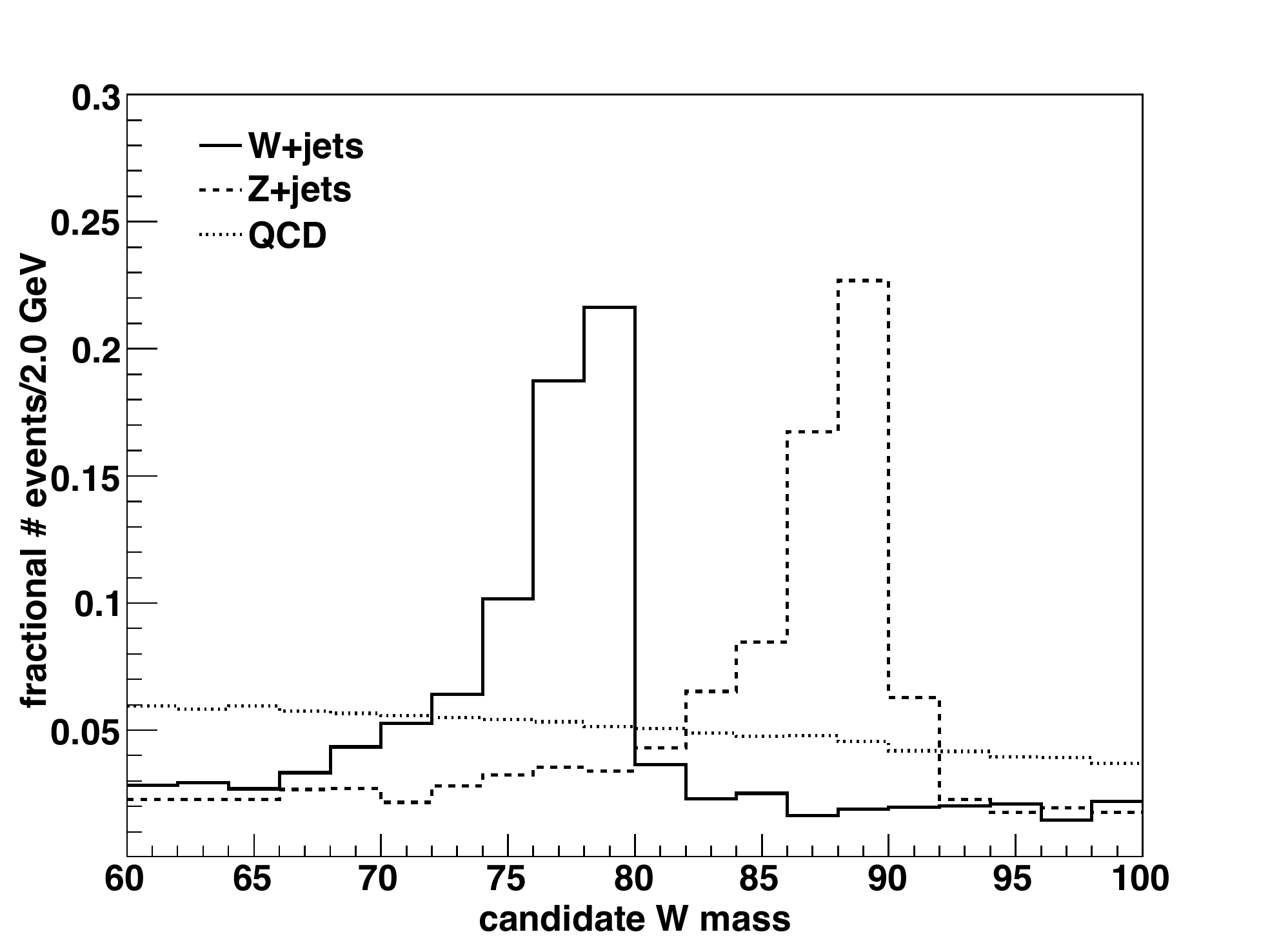}
\caption{Area normalized distribution of the jet mass for the $W/Z$-tagged jets when the tagger works on  $W + \jets$, $Z + \jets$ and QCD dijet samples.}
\label{fig:wtagged}
\end{figure}

\item {\bf Higgs tagger:}  The methodology of our Higgs tagger is identical to the BDRS algorithm as described in Ref.~\cite{Butterworth:2008iy}.  The tagger processes all tagged $b$-jets of size $R=1.2$ clustered with the Cambridge/Aachen algorithm.  The output of the tagger is a distribution of the jet substructure resonance mass, which contains both candidate Higgs bosons as well as backgrounds (from SM and new physics).  In this paper we present these distributions, as well as an estimated signal significance of Higgs discovery given a peak and mass window.  The full ``Higgs tagger'' is thus both extracting the distribution and specifying the candidate resonance jet mass window.  Any boosted massive particle that decays to $\bar{b} b$ is identified, and thus very good resolution of the candidate resonance jet mass observable is critical. Heavy-flavor identification is an essential ingredient for picking out particles that are consistent with Higgs bosons. Encouraged by the results of the study in Ref.~\cite{Plehn:2009rk,ATL-PHYS-PUB-2009-088}, our simulations assume a flat $70\%$ id efficiency for single $b$-tag with  $1\%$ fake rate~\footnote{Though we use an optimistic $b$-tagging efficiency, our approach is still somewhat conservative because we actually tag each Higgs candidate {\em three} times; one tag to select the jet for BDRS tagging, then two more tags on BDRS subjets.}.
\end{itemize}

\begin{figure}[t]
\centering
\includegraphics[width=0.48\textwidth]{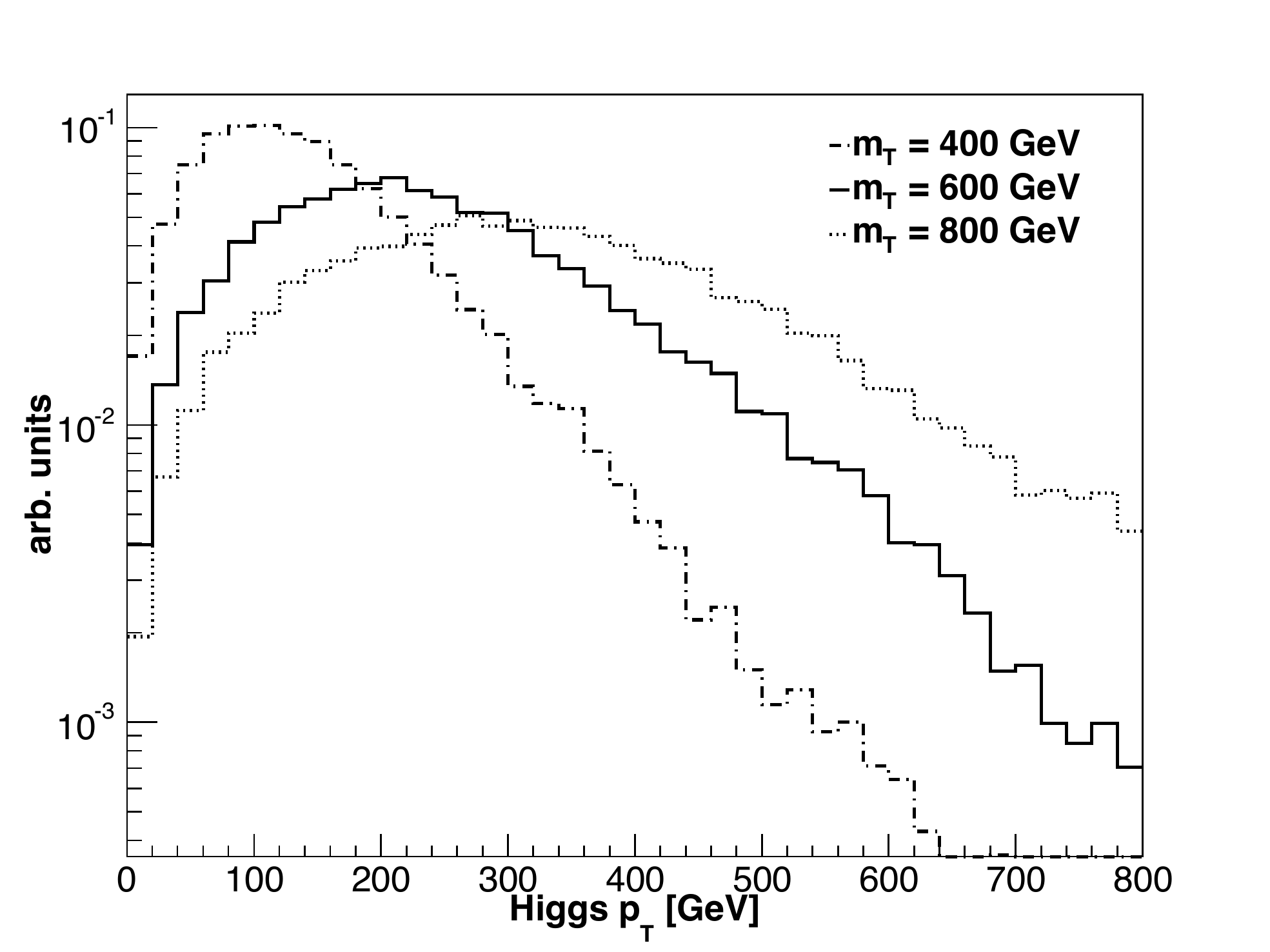}
\caption{Plot of the Higgs boson $p_T$ in $T$ decay for three different $T$ masses.}
\label{fig:higgspt}
\end{figure}

Given that top-partner pair production results in purely SM final states that are generically boosted, our search strategy has been optimized precisely to find the boosted fraction of the signal.  The boost distribution of the Higgs boson from the cascade decay of a top-partner is shown in Fig.~\ref{fig:higgspt}.  The Higgs boson $p_T$ peaks at roughly $m_T/2$, which means a large fraction of the events containing a Higgs boson are boosted for \emph{all} of the $T$ masses considered in this paper.  
The boosted jet substructure techniques are particularly useful to separate the Higgs boson decay to $b$-jets from $b$-jets that result from either decay of $t$ or $T$. Additional cuts on moderate/large $H_T$ and isolated lepton(s) further improves the signal significance.

As explained before, the Higgs tagger finds $b$-jets (would-be candidates for Higgs boson), but we employ it only after an input event passes through a series of selection criteria. We designed the criteria to capture all distinct signal topologies, rather than focusing on a subset. The different topologies are carefully chosen such that no event can be double-counted. In Fig.~\ref{fig:cutflow} all the channels of the selection procedure are depicted in a flow chart which we will briefly walk through now:
\begin{figure*}
\centering
\includegraphics[width=0.8\textwidth]{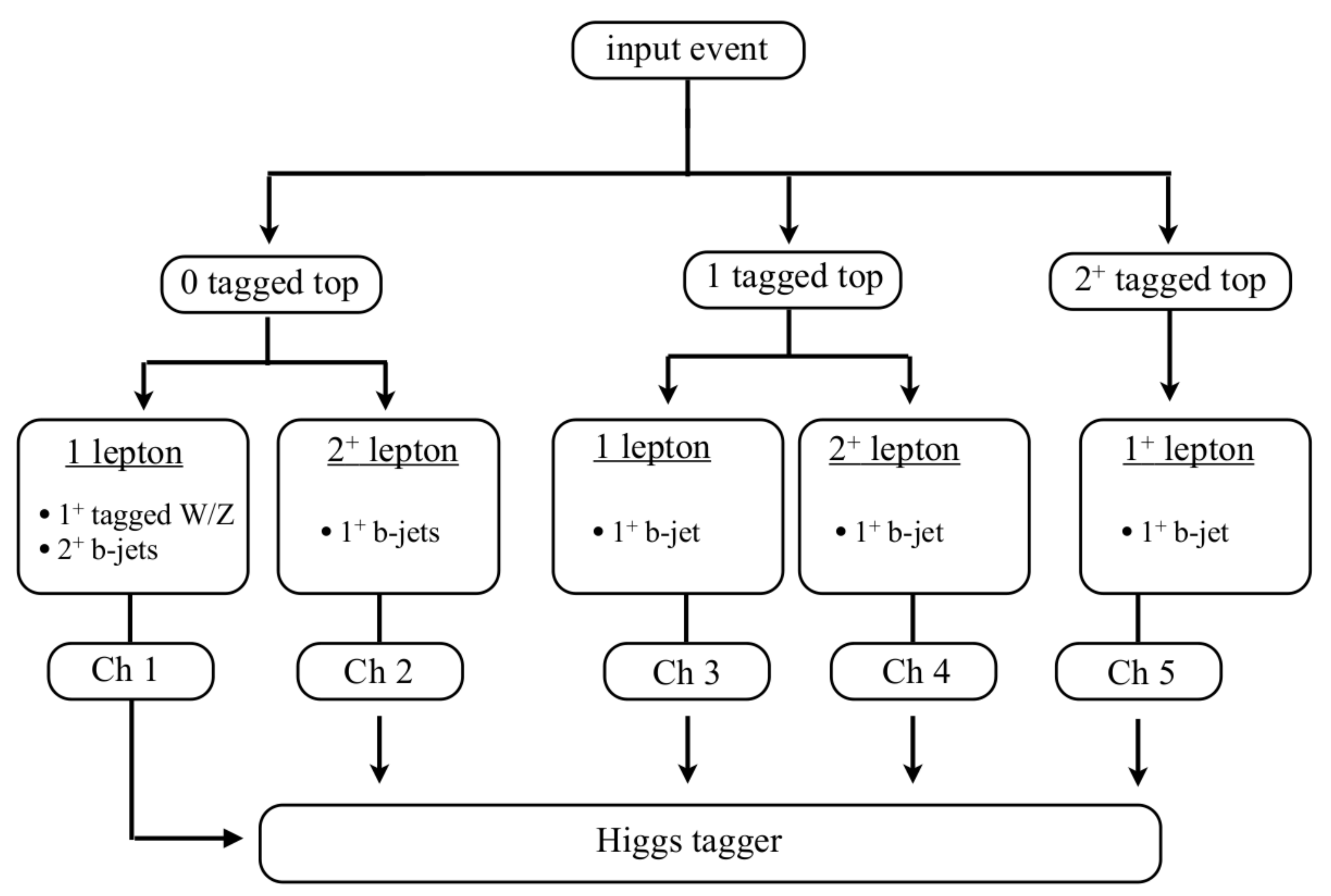}
\caption{Flow diagram to summarize our selection procedure. The flow has been devised so that no event can be counted more than once. In the figure, $1^+$ means one or more objects are required. Additional cuts on $H_T > 1,1.3\tev$ are imposed for top-partner masses of $800,1000\gev$ respectively.}
\label{fig:cutflow}
\end{figure*}
\begin{enumerate}
\item The top tagger runs on an input event and the event is classified in terms of the number of tops it identifies.  All the cells contained within tagged tops are removed from the event.  
\item The event is then further classified according to the number of isolated leptons present.
\item Events with large top and/or lepton multiplicities -- in particular, events with a single top $+$ more that one lepton  or two tops $+$ at least one lepton -- have little standard model contamination. We select these events immediately as long as they contain at least one additional $b$-jet on which the Higgs tagger can be run.  
\item Zero top $+$ a single lepton is the most contaminated channel. To extract a good signal/background we employ additional requirements:  the $W/Z$ tagger finds at least one $W/Z$ and it contains at least two $b$-jets, one of which contains a Higgs boson candidate. The additional $W/Z$ requirement and the extra $b$-tag are necessary to bring $W/Z +\ \jets$ under control. 
\item For the rest of the channels, with one top $+$ one lepton or no top $+$ two (or more) leptons we demand that there be at least one $b$-jet.
\end{enumerate}
After an event gets selected, the Higgs tagger runs over all leftover $b$-jets to determine whether these contain two $b$-tagged subjets consistent with the two-body decay of a Higgs boson.  Once a Higgs-tagged jet is identified, we plot the jet mass.  A peak in the signal events, above the $Z$ mass but below where $h \ra WW$ dominates, provides the evidence for a resonance consistent with a Higgs boson. 

Having outlined our analysis strategy, we now present the details of our signal and background simulations and summarize our results. 
 
\section{Simulation details and results}
\label{sect:sim}

To simulate the $T$ pair production we use Madgraph and Madevent~\cite{Maltoni:2002qb, Alwall:2007st}.  The production cross sections are given in Table~\ref{table:xsecsignal}.\vfil
\begin{centering}
\begin{table}[h!]
\centering
\begin{tabular}{|c|c|}
\hline
$m_{T}$ & $\sigma(pp \rightarrow T\bar T )$ \\ \hline
$400\ \gev$ & $12.7\pb$ \\ 
$600\ \gev$ & $1.29\pb $ \\ 
$800\ \gev$ & $0.229\pb$ \\ 
$1\ \tev$   & $0.054\pb$ \\ \hline
\end{tabular}
\caption{$T$ pair production cross section at a $\sqrt s = 14\tev$ center of mass LHC at NLO with MCFM.}
\label{table:xsecsignal}
\end{table}
\end{centering}
The heavy quark pair are subsequently decayed, showered and finally hadronized by PYTHIAv6.4~\cite{Sjostrand:2006za}.  For the signal events, we also use the ATLAS tune~\cite{Buttar:2004iy} in PYTHIA to model the underlying event. For the background, events are simulated at parton level using ALPGENv13~\cite{Mangano:2002ea}, then similarly showered and hadronized using PYTHIA.  

We do not use any realistic detector simulation or smearing in this work. The PYTHIA output events  are directly used as input to our analysis. In each event, isolated leptons (we neglect the possibility of jets faking leptons here), isolated photons, and neutrinos (invisible to the detector) are identified while the rest of the particles (out to $|\eta| \le 4$) are granularized into calorimeter `cells' of size $0.1 \times 0.1$ in $(\eta, \phi)$. The three-momentum of each scale is rescaled so that the cell is massless, and all cells of energy less than $1\gev$ are removed. The surviving cells are fed into the analysis chain described in Sec.~\ref{sect:search} above. Jet clustering is done using the inclusive Cambridge/Aachen algorithm, as implemented in Fastjet~\cite{Cacciari:2005hq}.

Once the events have been processed in the various channels, the final step in the Higgs search is to plot the Higgs candidate jet mass for all passing events and look for a feature. This can be done for each of the channels defined in Fig.~\ref{fig:cutflow} or by summing all channels together. The invariant mass of the Higgs candidates summed across all channels for $m_T = 400$, $600$, $800$ and $1000\gev$ are shown below in Fig.~\ref{fig:mt600}.

\begin{figure*}[t]
\begin{center}
\includegraphics[width=0.45\textwidth]{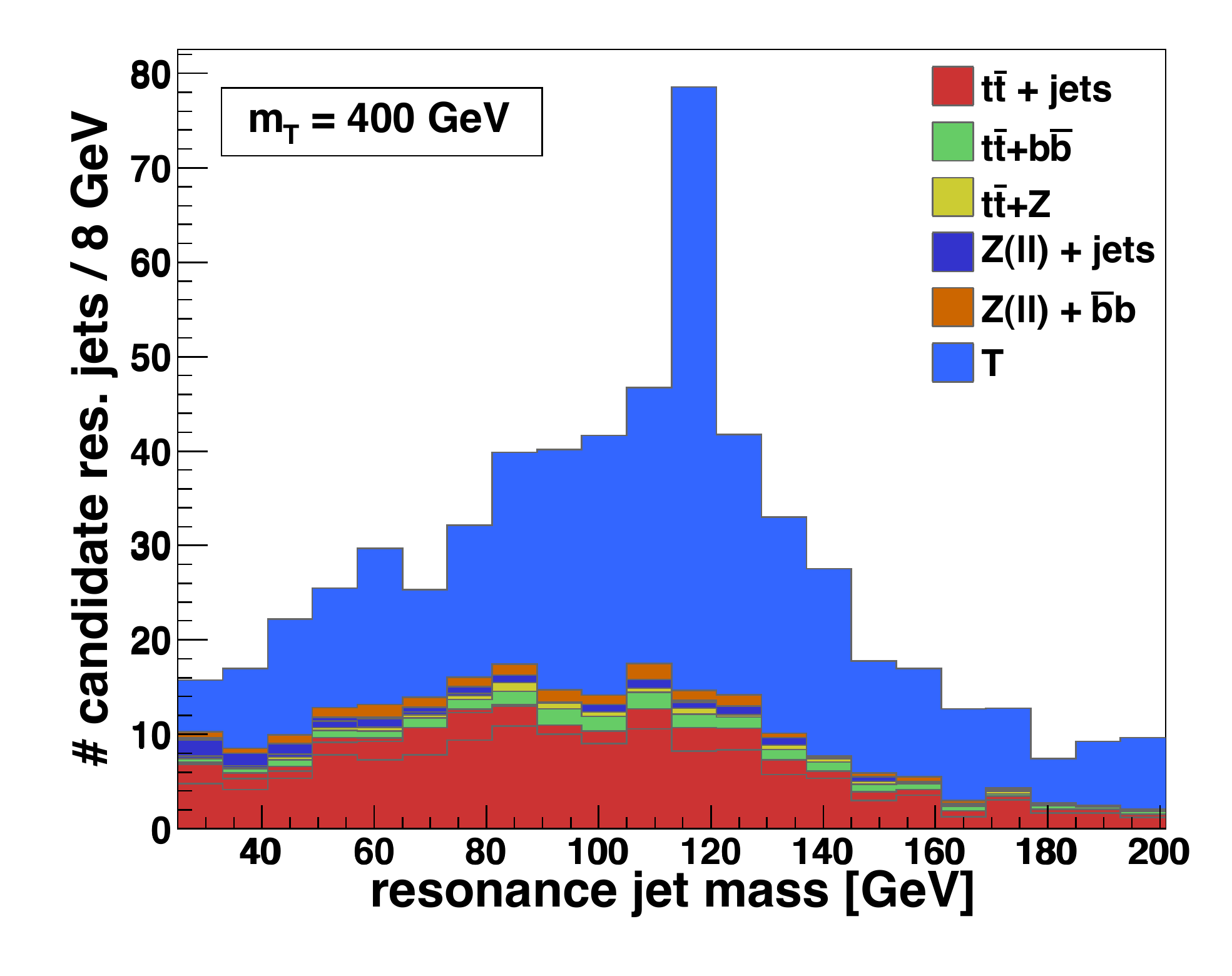} \hspace*{0.05\textwidth}
\includegraphics[width=0.45\textwidth]{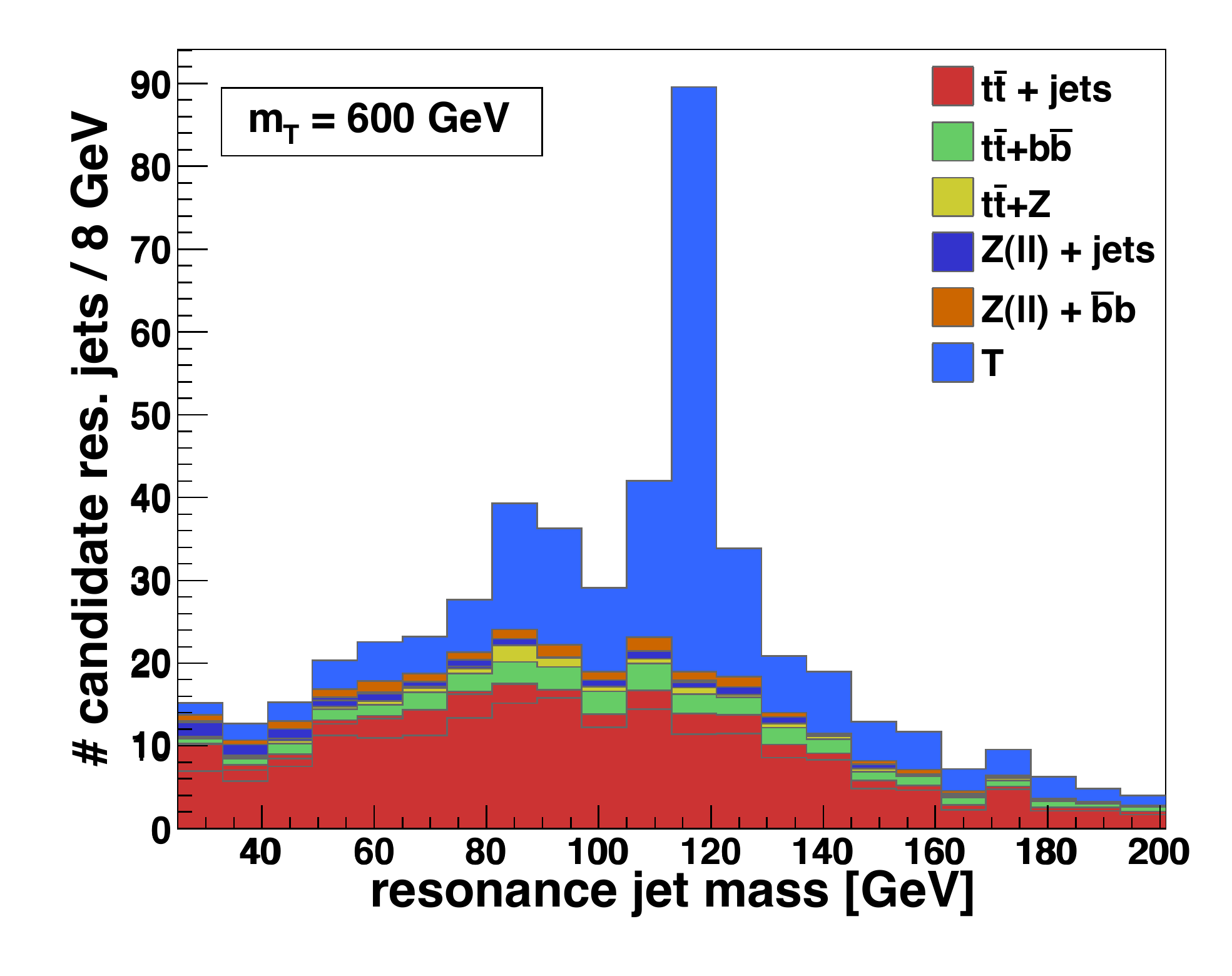} \\
\includegraphics[width=0.45\textwidth]{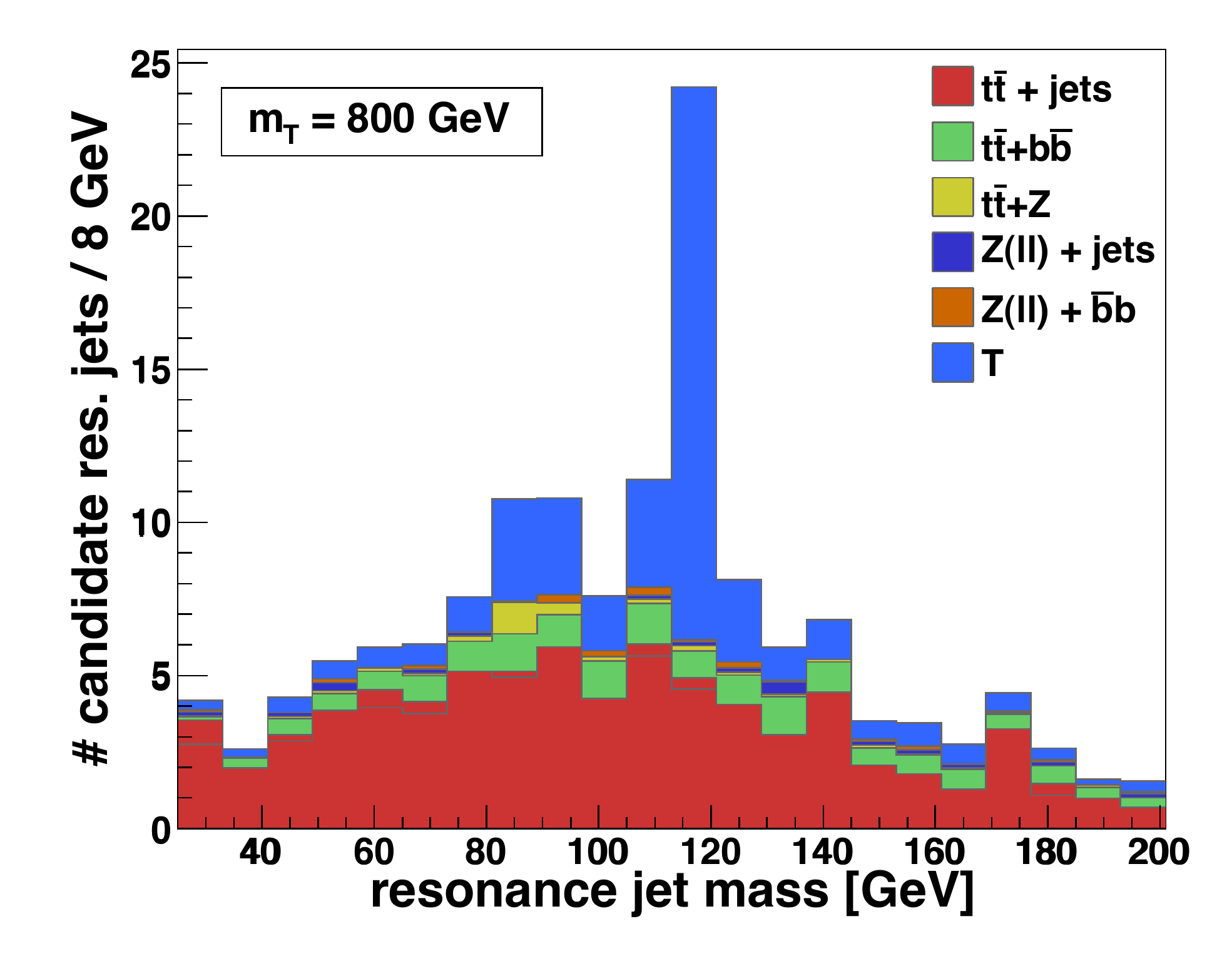} \hspace*{0.05\textwidth}
\includegraphics[width=0.45\textwidth]{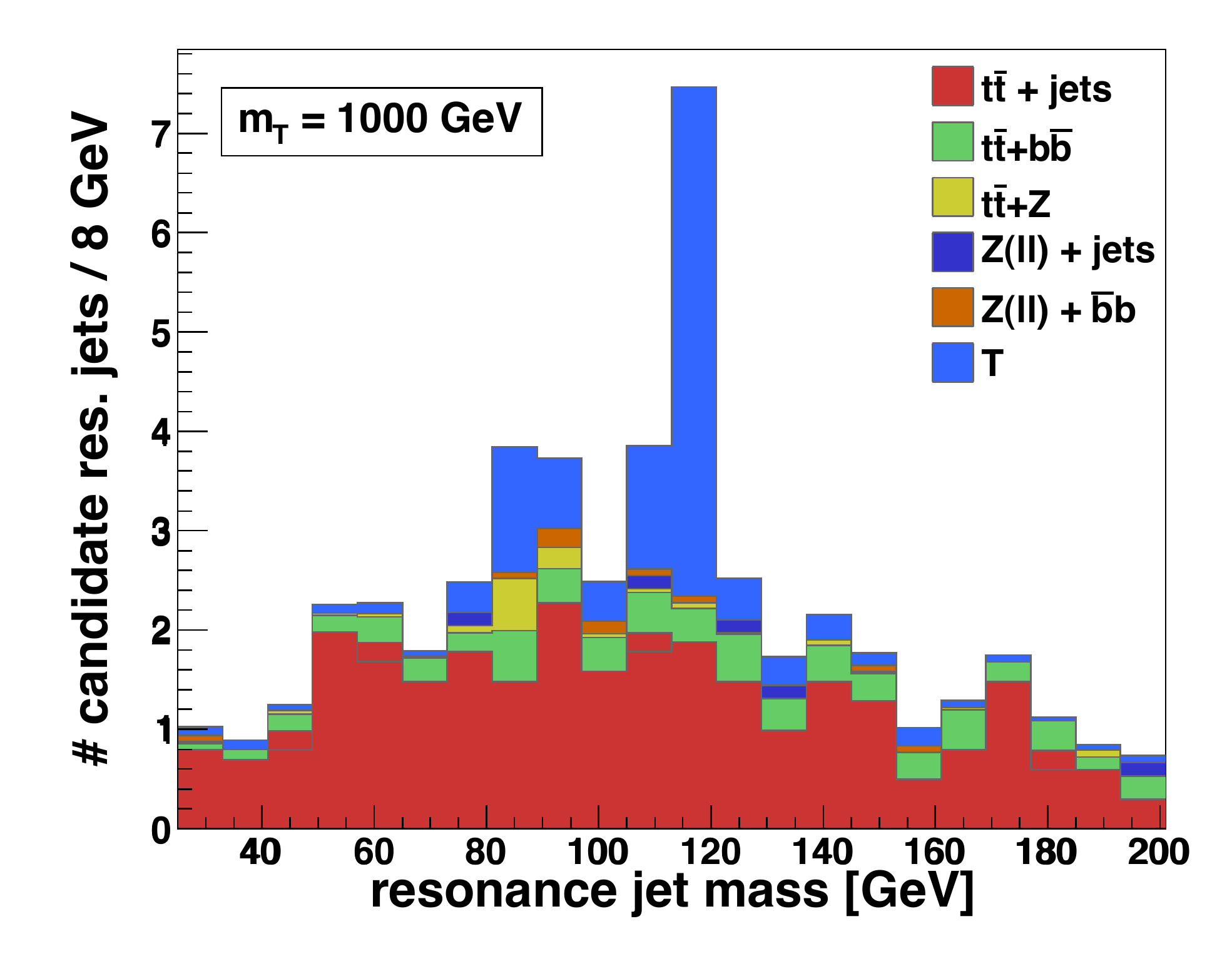} 
\end{center}
\caption{Resonance jet mass distribution. We assume an integrated luminosity of $10\fb^{-1}$ at a $14\tev$ center of mass LHC. The search strategy is described in Sec.~\ref{sect:search} and in Fig.~\ref{fig:cutflow}. An additional cut on $H_T > 1,1.3\tev$ is imposed for top-partner masses of $800,1000\gev$.}
\label{fig:mt600}
\end{figure*}

The peak in jet mass, consistent with Higgs decay, rises prominently above the background in all four examples. The new physics events to the left of the Higgs peak come predominantly from hadronically decaying $W/Z$. According to the branching ratios in Eq.~\eqref{eq:6}, equal amounts of Higgs bosons and $Z$ are produced in $T$ decay, however the branching fraction of $Z\rightarrow \bar b b$ is $\simeq 0.15$, so the $Z$ feature is small. Hadronic $W$ are also produced in $T$ decay, however the majority of them are removed by cuts preceding the Higgs-tagger.

% Hadronic $W$ are much more common than Higgs bosons. However, they do not pass the $b$-tagging criteria except when a light jet fakes a $b$, roughly a factor of $\sim 30$ suppression.

The measure we use for how well a particular analysis or sub-channel performs is the significance, which we define simply as $S/\sqrt B$. In addition to the SM background, we also include the new physics background --  hadronically decaying $W$, $Z$ from $T$ decay mentioned above. To account for this new physics background we define the total background to be the average of the number of events in the bins $\pm 2$ from the putative Higgs peak. This is a crude, and somewhat conservative estimate for the significance. The significance in each channel, as well as the significance of all channels combined is summarized below in Table~\ref{table:signifs}.

\begin{table}[t]
\centering
\begin{tabular}{| c || c|c || c|c || c|c || c|c ||}\hline
& \multicolumn{2}{|c||}{$400\gev$} 
& \multicolumn{2}{|c||}{$600\gev$} 
& \multicolumn{2}{|c||}{$800\gev$} 
& \multicolumn{2}{|c||}{$1\tev$} \\ \hline
& $S/\sqrt{B}$ & $S/B$ & 
  $S/\sqrt{B}$ & $S/B$ &
  $S/\sqrt{B}$ & $S/B$ &
  $S/\sqrt{B}$ & $S/B$ \\ \hline
Ch 1     &  2.0 & 0.4  &  4.3 & 1.3  &  2.0 & 1.0 & **  & **  \\ 
Ch 2     &  4.3 & 0.5  &  5.2 & 0.9  &  2.5 & 1.2 & **  & **  \\  
Ch 3     &   *  &   *  &  6.6 & 2.2  &  2.7 & 1.2 & 2.0 & 1.6 \\
Ch 4     &  2.7 & 0.7  &  4.4 & 1.8  &  **  & **  &  ** &  ** \\ 
Ch 5     &   *  &   *  &  4.1 & 1.1  &  3.1 & 1.2 & 1.4 & 0.8 \\ 
\hline
sum      &  5.2 & 0.5  & 10.5 & 1.2  &  5.2 & 1.2 & 2.4 & 1.2 \\ \hline
   \end{tabular}
\caption{The $S/\sqrt{B}$ and $S/B$ obtained for the various search channels, as well as for the summed significance of all channels.  The search was done for the heavy quark mass of $400\gev$, $600\gev$, $800\gev$ and $1\tev$, all assuming $\sqrt s  = 14\tev$ and $10\fb^{-1}$ of integrated luminosity.  The starred entries have significance less than 2 and are not included in the summed significance.  The double-starred entries have fewer than 2 events in the signal, and are also not included in the total significance.}
\label{table:signifs}
\end{table}

As Table~\ref{table:signifs} shows, the most significant channel(s) vary with the mass of the top-partner. When the top-partner is light, the cross section is large, but the boost of a Higgs or top from $T$ decay is smaller.  Consequently, channels which require several substructure tags (Ch.~$3, 5$) are inefficient, while the high cross section makes up for small branching ratios in the multi-lepton channels (Ch.~$2,4$). At high mass, the channels swap roles; the multi-lepton channels don't receive enough events, while the efficiency of substructure taggers improves greatly. For the intermediate point, $m_T = 600\gev$ all channels are equally effective. 

\section{Top-partners in Specific Models}
\label{sec:models}
Vector-like top-partners are  self-contained extensions of the standard model.  Nevertheless, they often appear as ingredients in larger extensions.  Here we demonstrate how our results from previous sections on jet substructure and Higgs-finding apply directly to two general classes of models: little Higgs theories and topcolor theories.  In particular, we map our parameter space onto two specific examples of these models:  the simplest little Higgs model~\cite{Schmaltz:2004de} and the top quark seesaw theory of electroweak symmetry breaking~\cite{Chivukula:1998wd}.

\subsection{The Simplest Little Higgs}

In the simplest little Higgs model, the Higgs boson is naturally light because it is a Nambu-Goldstone boson of a spontaneously broken symmetry  $(\SU(3)_W \times U(1)_X)/(\SU(2)_W \times U(1)_Y)$.  Nonzero vevs of two scalars (say, $\phi_1$ and $\phi_2$) in the triplet representation of $\SU(3)_W$ break the full symmetry down to $\SU(2)_W \times U(1)_Y$ at the scale $f > v$. Interactions of SM Higgs doublet can easily be calculated in the following parametrization of  the $\phi_i$ fields:
\begin{align}
  \label{eq:12}
  \phi_1 &= \exp \left\{ i \begin{pmatrix}  & H^\dag\\ H &  \end{pmatrix} \right\}
    \begin{pmatrix}  \\ f \end{pmatrix}  \\
  \phi_2 &= \exp \left\{- i \begin{pmatrix}   & H^\dag\\ H &  \end{pmatrix} \right\}
    \begin{pmatrix}  \\ f \end{pmatrix}  
\end{align}

The quadratic divergences associated with the top Yukawa is cancelled by extending  the $\SU(3)$ symmetry to the Yukawa couplings. First, the quark doublets are enlarged into $SU(3)$ triplets: $\Psi \equiv \left( Q_3, T\right)$, transforming under the $\SU(3)_W$ gauge symmetry. Second, two color-triplet, $SU(3)_W$-singlets $T^c_1$ and $T^c_2$ are introduced. The $U(1)_X$ charges of $T^c_1$ and $T^c_2$ are chosen to be equal and identical to the $U(1)_X$ charge of $t^c$.
\begin{equation}
  \label{eq:10}
  \mathcal{L}_\text{Yukawa} = \lambda_1  \phi^\dag_1 \Psi T^c_1 
              + \lambda_2  \phi^\dag_2 \Psi T^c_2  \; .
\end{equation}
Expanding $\phi_1$ and $\phi_2$ around their vevs
 (as in Eq.~\eqref{eq:12}) we find the effective Lagrangian 
\begin{equation}
  \label{eq:11}
  \begin{split}
     \mathcal{L} \supset & \  H Q_3 \left( \lambda_1 T_1^c  - \lambda_2 T_2^c \right)   \\
           & +  f T  \left( \lambda_1 T_1^c + \lambda_2 T_2^c \right) 
                   \left( 1 - \frac{1}{2 f^2} H^\dag H \right) + \dots  \; .
  \end{split}
\end{equation}
The above set of interactions can be recast in the form of Eq.~(\ref{eq:1}) by the following substitutions
\begin{equation}
  \label{eq:14}
  \begin{split}
     t_c  = \frac{1}{\sqrt{\lambda_1^2 + \lambda_2^2}} &
             \left(\lambda_1 T_1^c - \lambda_2 T_2^c \right)  \\
    T_c  = \frac{1}{\sqrt{\lambda_1^2 + \lambda_2^2}} &
             \left(\lambda_2 T_1^c + \lambda_1 T_2^c \right)  \\
    y_1 = & \sqrt{\lambda_1^2 + \lambda_2^2} \\
   \delta =  \frac{\lambda_1^2 - \lambda_2^2}
                  {\sqrt{\lambda_1^2 + \lambda_2^2}}\, f\;,  \quad &
  M  =  \frac{2 \lambda_1 \lambda_2}
                  {\sqrt{\lambda_1^2 + \lambda_2^2}}\, f\,.\nonumber
  \end{split}
\end{equation}

\subsection{Top quark seesaw theory}

In topcolor models, the Higgs boson is a composite particle formed 
from top quarks and new, heavier fermions as the result of some 
TeV-scale strong dynamics.  In the top quark seesaw 
theory \cite{Chivukula:1998wd}, the gauge structure is enlarged to
$SU(3)_1 \times SU(3)_2 \times SU(2)_W \times U(1)_Y$.
The third generation of SM quarks, along with an additional pair of quarks,
are embedded as
\begin{eqnarray*}
Q'({\bf 3},{\bf 1},{\bf 2},1/6), \quad T'({\bf 3},{\bf 1},{\bf 1},+2/3), \\
T'^c({\bf 1},{\bf \bar{3}},{\bf 1},-2/3), \quad t'^c({\bf 1},{\bf \bar{3}},{\bf 1},-2/3) .
\end{eqnarray*}
Topcolor is broken to QCD, 
$SU(3)_1 \times SU(3)_2 \ra SU(3)_C$,
through a scalar link field transforming as a 
$({\bf \bar{3}},{\bf 3},{\bf 1},0)$.
This results in a composite Higgs doublet formed from a bound state of
$Q'$ with $t'^c$.  The low energy effective theory
is just the SM group with an additional pair of quarks that transform under
a vector-like representation as well as a color octet
of massive gauge bosons.  Below the scale of the color octet,
the theory contains the identical particle content and interactions
given by Eq.~(\ref{eq:1}).  

Translating Ref.~\cite{Chivukula:1998wd} into the notation 
presented here requires the fields be relabeled as
\begin{eqnarray*}
t_L    &\ra& t_L     \\
\chi_L &\ra& T       \\
t_R    &\ra& {T^c}^* \\
\chi_R &\ra& {t^c}^*, 
\end{eqnarray*}
and the mass terms relabeled as
\begin{eqnarray*}
m_{t \chi}     &\ra& m   \\
\mu_{\chi t}   &\ra& M   \\
\mu_{\chi\chi} &\ra& \delta . 
\end{eqnarray*}
This ``schematic model'' involving only one composite Higgs boson
has a large Higgs mass, $\mathcal{O}(1 \; {\rm TeV})$.  
However, as emphasized by 
Ref.~\cite{Chivukula:1998wd}, a more general theory of the
the seesaw mechanism for electroweak symmetry breaking involves
more composite scalars, allowing one of the neutral Higgs bosons
to be as light as $\mathcal{O}(100 \; {\rm GeV})$.

One example of a more general topcolor theory involves 
extra dimensions \cite{Cheng:1999bg}.  In the extra dimensional form, 
a strongly interacting top quark is described by promoting the 
right-handed top into a 5D bulk fermion and giving it a profile which  
overlaps with the strong sector.  Reduced to a 4D theory, 
the KK modes of $t_R$ become fermionic resonances, 
more massive copies of $t_R$ with exactly the quantum numbers of the 
top-partner we are studying.  Boundary conditions slightly mix the 
various $t_R$ modes. 
When truncated to the zero modes of $t_L, t_R$ and the first $t_R$ 
KK excitation, the top quark mass matrix has exactly the form 
of our Eq.~(\ref{eq:2}).  Other extra dimensional models can be found
in Ref.~\cite{Carena:2006bn, Contino:2006qr,Burdman:2007sx} and
deconstructed versions in Ref.~\cite{Chivukula:2009ck}.

\section{Discussion}
\label{sec:conclude}

We have demonstrated that top-partner production and decay leads 
to highly boosted, light Higgs bosons that can be discovered in their 
dominant $\bar b b$ decay mode through the use of jet substructure 
techniques.  For top-parters with masses in the range $400-800$ GeV,
we explicitly showed that our estimate of the $S/\sqrt{B}$ 
of Higgs discovery exceeds $5$ with just $10\fb^{-1}$ of $14\tev$ LHC data.  
This is all the more remarkable given that top-partner decay
results in just SM particles in the final state.  We demonstrated that
a multi-channel analysis yields the best approach to maximizing
Higgs boson discovery significance.  
Our analysis thus explicitly demonstrates the full extent of our 
original observation in 2009 \cite{Kribs:2009yh} that Higgs production 
and decay to $b\bar{b}$ through new physics production may well 
be \emph{the} discovery mode of the Higgs boson.
Finally, the methods we employed in the paper to find the Higgs boson
can also be applied as additional search channels for 
top-partner production itself.  
Comparing the significance of our channels against older studies
\cite{Skiba:2007fw, Holdom:2007nw} suggests that including the
Higgs boson in the final state may well accelerate the discovery
of top-partners!

It is interesting to compare and constrast top-partner production and
decay against the superpartner production and decay to Higgs bosons.
One of the main ingredients of \cite{Kribs:2009yh,Kribs:2010hp} 
was to isolate the supersymmetric signal from SM backgrounds 
using large missing energy, which resulted from the escaping
lightest supersymmetric partners.  Top-partner production and decay,
by constrast, always results in just SM particles in the final state.
The reason top-partner production is \emph{viable} is mainly
because of the large fraction of top-partner decays that contain 
a substantially boosted Higgs boson.  
We demonstrated ({\it c.f.}\ Fig.~\ref{fig:higgspt}) that the Higgs boson 
$p_T$ peaks at roughly $m_T/2$, whereas the Higgs boson typically
receives less of a boosted in a supersymmetric cascade, 
$\mathcal O(m_{\tilde q}/4)$, and thus requires more massive
squarks, which lowers the production cross section.
In supersymmetry, this is partially compensated by the
potentially large number of squark production channels that
could lead to Higgs bosons in the final state \cite{Kribs:2010hp}.

Finally, while we have focused on the top-partner in isolation, 
there are many models for which it is but one component of a 
larger extension beyond the standard model.  We explicitly
demonstrated that the particle content and interactions
arise precisely in the simplest little Higgs model \cite{Schmaltz:2004de}
as well as the top quark seesaw model of electroweak symmetry 
breaking \cite{Chivukula:1998wd}.  These are but two examples
of a large class of models where, if the Higgs boson is light enough,
it can be discovered through new physics production and decay.
The case of topcolor models is particularly interesting, since they
also have a massive color-octet vector boson that couples
to (light and heavy) quarks, potentially dramatically increasing the 
production cross section of top partners if the color-octet is not 
too heavy \cite{Dobrescu:2009vz}.  Bottom-partners also provide
an interesting accompanying particles that can result in 
Higgs bosons.  However, since bottom-partners could decay to 
$bh, bZ$ and $Wt$, in complete analogy to $T$,  they can pollute $T$ events 
making the extraction of the Higgs boson harder.  However, 
a nearly degenerate bottom-partner can also 
provide great Higgs discovery opportunities on its own.
More elaborate vector-like extensions of the SM also exist where
the branching fraction to $h + X$ can increase dramatically to even 
$100\%$~\cite{Martin:2009bg} making this analysis even more significant. 
However, the extreme branching fractions require special combinations 
of field content and mixing angles.

\section*{Acknowledgments}

We thank B.~Dobrescu for helpful discussions.
This work was supported in part by the US Department of Energy 
under contract numbers DE-FG02-96ER40969 (GDK, TSR) and 
DE-FGO2-96ER40956 (TSR). 
GDK was supported by a Ben Lee Fellowship from Fermilab.
AM is supported by Fermilab operated by Fermi Research Alliance, 
LLC under contract number DE-AC02-07CH11359 with the 
US Department of Energy. \\

\appendix*

\section{Background details}

In  Table~\ref{table:xsec} we  summarize all the background events we have considered in this work. We list their cross sections along with the  the  parton-level cuts we use to generate these events. To avoid over-counting in the  $\bar t t+\ \jets$ and $W/Z + \ \jets$ backgrounds, MLM jet-parton matching was performed according to the procedure outlined in~\cite{MLM,Catani:2001cc,Alwall:2007fs}. 
\begin{centering}
\begin{table}[h!]
\centering
\begin{tabular}{|c|c|}
\hline
Process & $\sigma_{LHC}$\\ \hline
$\bar t t + 0\ \jets$ & $254\pb $ \\ 
$\bar t t + 1\ \jets$ & $133\pb$ \\
$\bar t t + 2^+\ \jets$ & $71\pb$ \\ \hline
$\bar t t + \bar b b$ & $2.6\pb$ \\ \hline
$\bar t t + Z$ & $1.1\pb$ \\ \hline
$Z(\ell\ell) + 2\ \jets$ & $80\pb$  \\ 
$Z(\ell\ell) + 3^+\ \jets$ & $29\pb$ \\ 
$Z(\ell\ell) + \bar b b$ & $82\pb$ \\
$Z(\ell\ell) + \bar b b + \jet$ & $31\pb$ \\ \hline 
\end{tabular}
\caption{Background cross sections at a $\sqrt s = 14\tev$ center of mass LHC. CTEQ5L pdfs and default renormalization and factorization scales were used for all background processes. Parton level cuts of  $p_{T, j} > 25\gev, |\eta_j| < 4, \Delta R_{jj} > 0.4$ were applied when generating all events with the exception that no $p_{T}$ or $|\eta|$ requirements were placed on the $b$-jets from $W/Z + \bar b b$. The $\bar t t + \jets$, $\bar t t + \bar b b$ and $\bar t t + Z$ cross sections have all been scaled to NLO using $K$ factors taken from~\cite{Plehn:2009rk, Bevilacqua:2010ve, Dittmaier:2008uj, Bevilacqua:2009zn, Bredenstein:2009aj, Lazopoulos:2008de}.}
\label{table:xsec}
\end{table}
\end{centering}

In addition to the above backgrounds, we checked $W\,+\,2,\, 3\ \jets$ and $W + \bar b b\, +\, 0,1\ \jets$. These processes, especially $W+ \jets$, have large cross sections, but they fall mostly within the $1$ lepton, $0$-top channel. By requiring an extra $b$-jet and an extra tagged $W/Z$ these backgrounds can be successfully mitigated. We find they comprise a few percent, at most of the background above, so we do not include them in our analysis. 

\bibliography{boosted_TReference}

\end{document}